\documentclass[onecolumn]{IEEEtran}
\usepackage{amsmath,amssymb}
\usepackage{graphicx}
\usepackage[justification=centering]{caption}
\usepackage{mathrsfs,amssymb}
\usepackage{longtable}
\usepackage{mathtools} 
\usepackage{epstopdf}
\usepackage{dsfont}
\usepackage{amsthm}
\usepackage{cite}
\usepackage{setspace}
\usepackage{color}
\usepackage{dblfloatfix}
\newcounter{Mytempeq}
\makeatletter
\def\th@plain{%
\thm@notefont{}
  \itshape 
  }
\def\th@definition{%
\thm@notefont{}
  \normalfont 
  }
\makeatother

\begin{document}
\title{ \ \\ Impact of User Height on the Coverage of 3D Beamforming-Enabled Massive MIMO Systems}
\author{\IEEEauthorblockN{Mahdi Baianifar$^\text{1}$, Soheil Khavari$^\text{1}$, S. Mohammad Razavizadeh$^\text{1}$} and Tommy Svensson$^\text{2}$\\
$^\text{1}$Iran University of Science and Technology (IUST), Iran \\
$^\text{2}$Chalmers University of Technology, Sweden  \\
\{mahdi{\_}baianifar,khavari\}  @elec.iust.ac.ir, rseyed@chalmers.se, tommy.svensson@chalmers.se}
\maketitle
\begin{abstract}
In this paper, we perform a coverage analysis of a cellular massive multiple-input multiple-output (MaMIMO) network which adopts 3D beamforming. In contrast to the previous works on 3D beamforming which assume that all users are placed on the ground (i.e., in a 2D environment), we consider a more practical scenario where the users of the network are dropped in a 3D environment  with different heights. In this scenario, by adopting a stochastic geometry framework, first we calculate the coverage probability of the network as a function of the users height. Then considering this coverage probability as the objective function, the optimum tilt angle of the base stations antenna pattern is found and the effect of users' heights on this tilt angle is investigated. Our numerical results show that by taking the users' heights into account and using the proposed method, a considerable improvement is achieved in the performance of the network compared to other similar 3D beamforming methods that ignore the users' height distribution. 

\end{abstract}
\IEEEpeerreviewmaketitle
\section{Introduction}

Massive multiple-input multiple-output (MaMIMO) in which a large number of antennas are deployed at base stations (BSs) of cellular networks has been proposed as one of the solutions for responding to the increasing capacity demands in 5G networks \cite{Marzetta_2010}. Providing high throughput, simple linear signal processing techniques,  reducing interference, averaging out the small scale fading, removing uncorrelated noise and reducing the transmit power are among the main advantages of MaMIMO systems. However, in practice, MaMIMO systems have some problems including hardware complexities and the pilot contamination effect \cite{EEandSEEngo13}. 

One of the related emerging techniques that is proposed for 5G networks is 3D beamforming (3DBF) which is feasible if we use a large number of antennas at the BS. The 3dBF can be applied both in analog beamforming (which is also known as 3d beam steering) and digital beamforming (which is known as precoding). In the analog 3DBF,  the antenna radiation pattern of a cellular BS antenna array is adjusted in both the horizontal and the vertical planes. Therefore, by using this technique, the BS can efficiently transmit/receive signals to/from desired directions and at the same time reduce the inter-cell or intra-cell interferences from unwanted sources. This brings a significant improvement in the network performance compared to the conventional 2D beamforming (2DBF) methods which ignore adaptation of the tilt angle of the BS \cite{DrRazav14}.  

In practice, duo to large beamwidth of the BS antennas radiation pattern in the horizontal plane, the 3DBF methods are mainly focusing on vertical beamforming i.e. adjusting tilt angle of the BS's antenna array \cite{CellCovOpt15,NetSumOpt16,3DMaMIMOMOd15, AdapMulticell16,DrRazav14,DLVBFinkyo14,SoheilPaper}. For example, by considering the tilt angle of the BS antenna radiation pattern, the cell coverage of a multi-cell MaMIMO network is investigated in  \cite{CellCovOpt15}. In \cite{NetSumOpt16}, a method for joint tilt angle optimization  and  transmit power allocation to maximize the sum spectral efficiency (SE) of a multi-cell MaMIMO network is proposed.
A joint precoding and antenna tilt optimization for maximizing energy efficiency (EE) is also proposed in \cite{SoheilPaper}.

In all the above mentioned papers and many other related research on the 3DBF, during optimizing the tilt angle of the BS antenna pattern, all users in a cell are assumed to be on the ground. More precisely, in those papers, the user terminals are  assumed to be distributed in a 2D environment with zero height or in a fixed height (typically 1.5 m). Hence, the effect of different users' height is ignored in the 3DBF design. However, this assumption is not true in real cellular networks in which the users can be in high buildings or in other words, have different heights. In these scenarios, modeling of the users' distribution in a 3D environment is closer to real applications. To the best of our knowledge, the only paper that considers users height in the 3DBF problem is \cite{DLVBFinkyo14} where the authors study a single user single cell network. 

Motivated by the above facts, in this work we revisit the problem of the 3DBF in a multi-cell MaMIMO network, by assuming that the users are dropped in a 3D environment  with different heights and investigate the effect of the users' height on the optimum tilt angles of the BSs antennas. First, we calculate the coverage probability in the uplink of a multi-cell massive MIMO network that uses the 3DBF and in addition users are distributed randomly in different heights. Moreover, the effect of the pilot contamination is also considered. To make our analyses more tractable and closer to practical random cellular networks, we adopt a stochastic geometry (SG) approach \cite{AtrackApp11} for calculating the coverage probability. In this way, the locations of the BSs are modeled based on a homogeneous Poisson point process (PPP). In addition, it is assumed that the users are distributed randomly in a 3D environment. 

Next, by using the calculated coverage probability as the objective function, the optimum tilt angle of the BSs antenna pattern is found. Our simulation results show that considering user's height in the analysis brings a significant performance gain compared to the methods that ignore users' height and only design the 3DBF method based on the location of the users on the x-y plane.

The rest of the paper is organized as follows: in Sec.~\ref{secsysmod} the proposed model for the system is described and in Sec.~\ref{seccov} the calculation of the coverage probability is presented. The tilt angle optimization problem is considered in Sec.~\ref{covmax}. Numerical results are presented in Sec.~\ref{simsec} and finally, Sec.~\ref{concsec} concludes the paper.
\vspace{-5pt} 
\begin{figure}
\centering
\includegraphics[scale = 0.25]{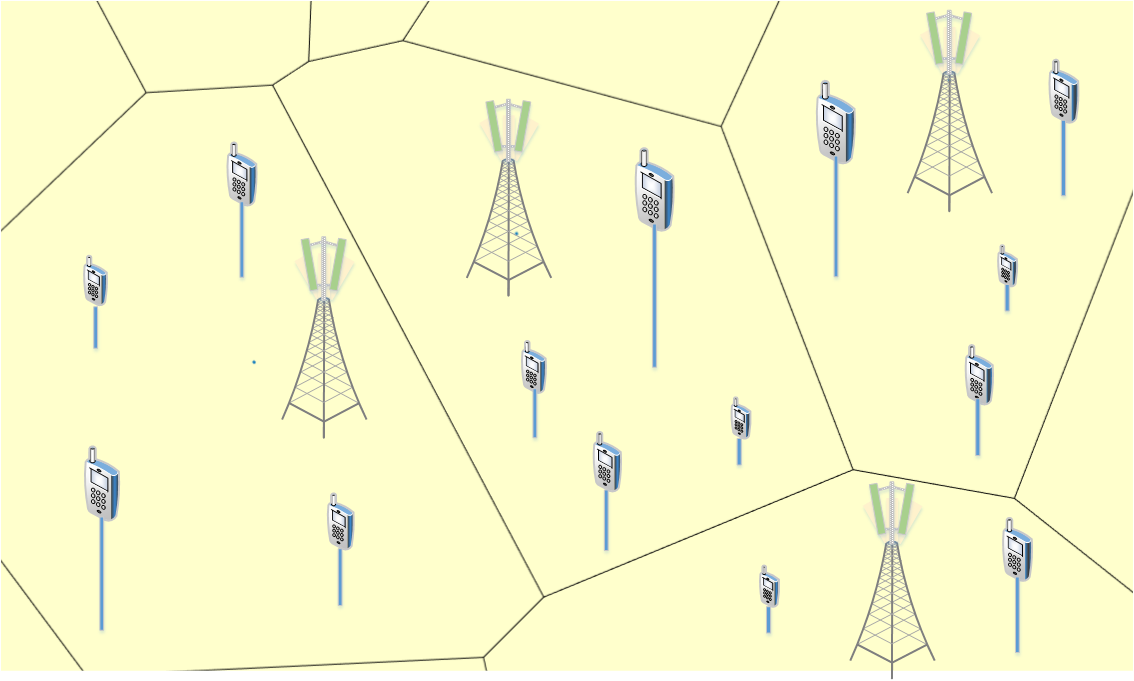}
\caption{Model of the cellular network with random BS deployment.} \label{FigSysMod}
\vspace{-25pt} 
\end{figure}

\section{System model} \label{secsysmod}
\begin{figure}
\centering
\includegraphics[scale = 0.235]{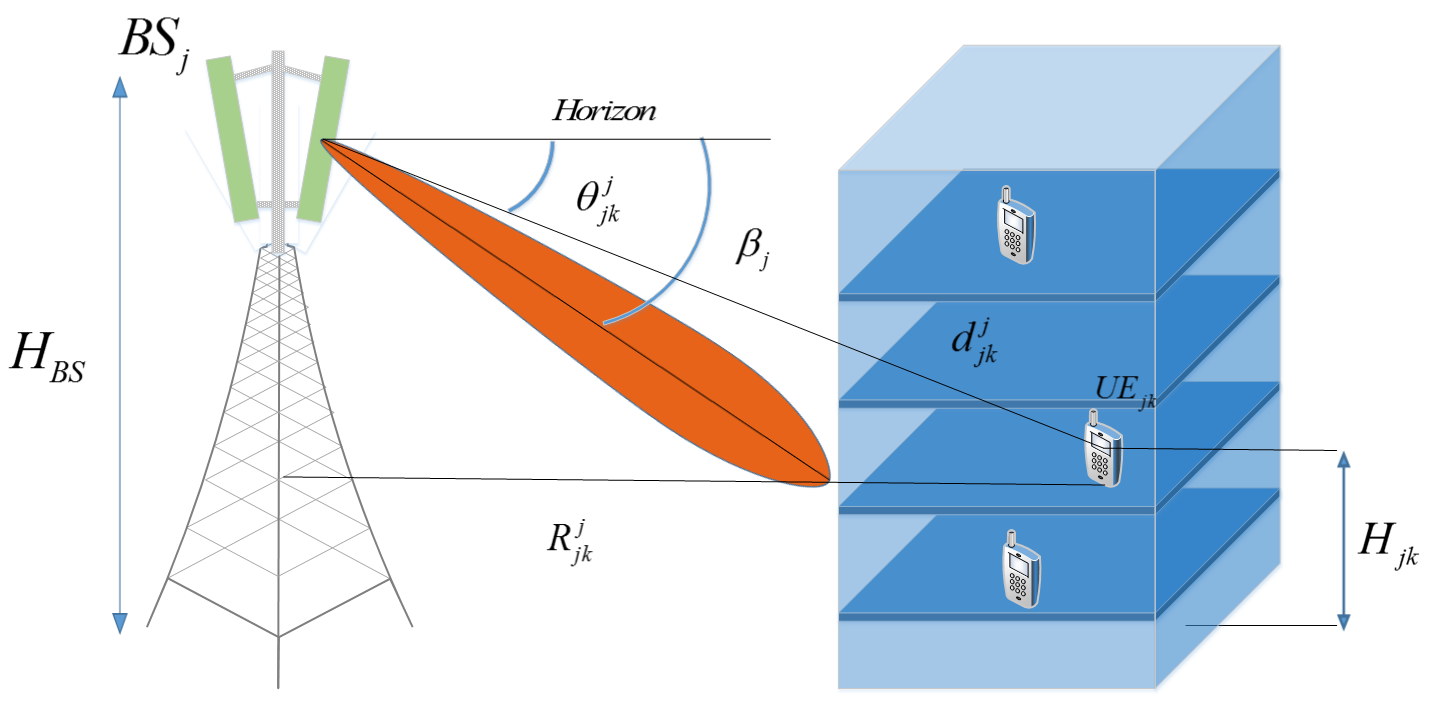}
\caption{Definition of the parameters in the system model.} \label{FigSysMod2}
\end{figure}

In our work, we consider the uplink of a cellular network in which the location of the BSs are drawn from a homogeneous PPP,~$\Phi$, in $\mathbb{R}^2$ with density $\lambda_m$ (see Fig.~\ref{FigSysMod}). 
Each BS is equipped with a very large number of antennas and serves $K$ users, simultaneously. The $k$'th user in the $j$'th cell is denoted by $\texttt{UE}_{jk}$ ($k=1,2,...,K$) and the $i$'th BS is denoted by $\texttt{BS}^i$. The vertical gain of the $\texttt{BS}^i$'s antenna radiation pattern in direction of the $\texttt{UE}_{jk}$ is denoted by $\alpha(\beta_i , \theta^i_{jk})$ and expressed as \cite{3GPPstd}
\begin{equation} \label{Tilt_Angle_Equation}
\alpha(\beta_i , \theta^i_{jk}) = 10^{-0.1 \times \min{\left[12 \left( \frac{\theta^i_{jk} - \beta_i}{\theta_{3dB}} \right)^2 , \ \text{SLL}_{el} \right]}}
\end{equation}
where as illustrated in Fig.~\ref{FigSysMod2}, $\beta_i$  is the tilt angle of the $\texttt{BS}^i$ antenna radiation pattern (i.e., the angle between the horizon and the main lobe of the radiation pattern) and $\theta^i_{jk}$ denotes the vertical user's angle of arrival (i.e., angle between the horizon and the line that connects $\texttt{UE}_{jk}$ to the $\texttt{BS}^i$). Also $\theta_{3dB}$ and $\text{SLL}_{el}$ are the half-power beamwidth (HPBW)  and the side lobe level of the $\texttt{BS}^i$ antenna radiation pattern in the vertical domain, respectively. 
As we see in Fig.~\ref{FigSysMod2}, $\theta^i_{jk} = \tan^{-1}({H^{eff}_{jk}} / {R^i_{jk}})$, where $H^{eff}_{jk} = H_{BS}-H_{jk}$ denotes the effective height of the BS as perceived  by the $\texttt{UE}_{jk}$. $H_{BS}$ and $H_{jk}$  are the height of the BSs and the $\texttt{UE}_{jk}$, respectively, and the horizontal distance between the $\texttt{UE}_{jk}$ and the $\texttt{BS}^i$ is denoted by $R^i_{jk}$. Using the above notations, the received signal at the $\texttt{BS}^m$ is
\begin{align} \label{ReceivedSig}
 \mathbf{y}_m & = \sum_{i=1}^K{\sqrt{\alpha(\beta_m , \theta^m_{mi}) L\left( d^m_{mi} \right)} \mathbf{h}^m_{mi}x_{mi}} \nonumber \\
 & + \sum_{\substack{\{ l: l \neq m,  Y_l \in \Phi\}}}{\sum_{i=1}^K{\sqrt{\alpha(\beta_m , \theta^m_{li}) L(d^m_{li})} \mathbf{h}^m_{li} x_{li}}} + \mathbf{n}_{m}
\end{align}
where $L( d^i_{jk})$, $\mathbf{h}^i_{jk}$ and $d^i_{jk}$  are the path loss, small scale fading vector and the distance between the $\texttt{UE}_{jk}$ and the $\texttt{BS}^i$, respectively. The path loss is obtained as  $L( d^i_{jk}) = C \cdot (d^i_{jk})^{-v}$, where $C$ is a scaling factor and $v$ is the path loss exponent. Also, $x_{jk}$ is the transmit signal from the $\texttt{UE}_{jk}$ and it is assumed that $E \left\lbrace \lvert x_{jk} \rvert^2 \right\rbrace = 1$. Finally, $Y_l$ shows the locations of the $\texttt{BS}^l$.

We assume that the UEs are located in different heights ranging from 1.5\text{m} for the users on the ground to 22\text{m} for the users located on a tall building. The BS height, $H_{BS}$, is also set to 32m as in \cite{3GPPstd}. The effective height $H^{eff}$ is between 10m and 30.5m and distributed according to a linear distribution as  
\begin{equation} \label{HeffEq}
f_{H^{eff}}\left(h \right) = a\left( b.h + c\right) + \left(1-a \right) \delta \left( h-30.5 \right)
\end{equation}
which is a slightly modified version of the distribution in \cite{DLVBFinkyo14} to taking into account the portion of users located on the ground. In this model, the parameter  $0\le a\le 1$ denotes the fraction of the users that are located on the ground and $b$ and $c$ are two constants depending on the environment and the height of the users. 

\begin{figure*}[!b]
	\normalsize
	\setcounter{Mytempeq}{\value{equation}}
	\setcounter{equation}{9}
	\hrulefill
	\begin{equation} \label{coverage_probability}
	P_c = \sum_{n=1}^N \left(-1 \right)^{n+1} \int_0^\infty{\exp{\left(-2\pi \lambda \int_{R_e}^{\infty}{r \left(1-E_{H^{eff}}\left\{    \exp{\left( - \eta \tau n{\frac{G_l}{G_0} \left( \frac{\sqrt{r^2 + h^2}}{\sqrt{x^2 + h_0^2}} \right)^{-v}} \right)} \right\} \right)}dr \right)}} f_R \left( x \right) dx
	\end{equation}
	\setcounter{equation}{\value{Mytempeq}}
\end{figure*}
According to the stochastic geometry approach and using the Slivnyak theorem \cite{AtrackApp11}, we study the performance of a typical user. Hereafter, we use index $0$ for the typical user and its serving BS. In the asymptotic case (i.e. MaMIMO) and  by considering the pilot contamination effect, the received SIR of the typical user at its serving BS can be expressed as
\begin{equation} \label{SINReq}
\text{SIR}  = \frac{\alpha(\beta_0 , \theta^0_{00}) L(d^0_{00})}{\sum_{\substack{\{l: l\neq 0,  Y_l \in \mathcal{N}_u^0 \}}}{\alpha(\beta_0 , \theta^0_{l0}) L(d^0_{l0})}}  
\end{equation}
where $\mathcal{N}_u^0$ denotes the set of the positions of all interfering users in other cells that use the same pilot as the typical user. In (\ref{SINReq}), an asymptotic case is considered in which noise and small scale fading are averaged out and the denominator is the coherent interference due to the pilot contamination effect \cite{Marzetta_2010}. In addition, the SIR in  (\ref{SINReq}) depends on the users' heights as well as the tilt angle of the BS antenna. In the next section, by using this SIR, we derive the coverage probability of the network. 

\section{The Coverage Probability}\label{seccov}
The coverage probability of the network is defined as 
\begin{equation} \label{cov1}
 \text{P}_c  = \text{Pr}{\left\{ \text{SIR}> \tau \right\} } 
\end{equation}
where $\tau$ is the SIR threshold value. It should be noted that the coverage definition in \eqref{cov1} is in fact the SINR coverage of the network which, as explained in Sec.~\ref{secsysmod}, in the  massive MIMO systems and in the asymptotic regime of an infinite number of antennas at all the BSs, equals to the SIR coverage. In addition, to model $\mathcal{N}_u^0$ in  \eqref{SINReq} (i.e. the location of the users that generate pilot contamination), some approaches have been proposed e.g. in \cite{pilotmodel_2015,UPLinkMMIMO16}. In this paper, we choose the solution in \cite{UPLinkMMIMO16} in which $\mathcal{N}_u^0$ is approximated as a homogeneous PPP with the same density as the BSs but with excluding users that are inside a ball centered at the $\texttt{BS}^o$ and with radius $R_e$. In this way, only users that have the distance larger than $R_e$ from the $\texttt{BS}^o$ and use the same pilot as the typical user are considered as the source of the pilot contamination for the typical user.

By considering this model, the coverage probability of the typical user can be derived as follows. By substituting \eqref{SINReq} in \eqref{cov1}, we have
{\small{
\begin{align} \label{SINRcaleq1}
& \text{P}_c  = Pr{\left\{\frac{\alpha(\beta_0 , \theta_{00}^0) L \left( d_{00}^0 \right)}{ \sum_{\substack{\{l: l\neq 0,  Y_l \in \mathcal{N}_u^0 \}}}{\alpha \left( \beta_0 , \theta_{l0}^0 \right) L \left( d_{l0}^0 \right)}} > \tau \right\} } = \nonumber \\
&\int_0^{\infty}{\text{Pr}{\left\{ \frac{\alpha \left( \beta_0 , \theta_{00}^0 \right) L \left( d_{00}^0 \right)}{ \sum_{\substack{\{l: l\neq 0,  Y_l \in \mathcal{N}_u^0 \}}}{\alpha \left( \beta_0 , \theta_{l0}^0 \right) L \left( d_{l0}^0 \right)}} > \tau | R^0_{00} = x \right\}} f_{R_{00}^0} \left( x \right) dx}
\end{align}}}
By denoting $G_0 \stackrel{\Delta}{=} \alpha ( \beta_0 , \theta^0_{00})$, $G_l \stackrel{\Delta}{=} \alpha ( \beta_0 , \theta^0_{l0})$, the conditional probability in (\ref{SINRcaleq1}) is obtained as
{\small{
\begin{align} 
&\text{Pr}{\left\{ \frac{\alpha \left( \beta_0 , \theta_{00}^0 \right) L \left( d_{00}^0 \right)}{ \sum_{\substack{\{l: l\neq 0,  Y_l \in \mathcal{N}_u^0 \}}}{\alpha \left( \beta_0 , \theta_{l0}^0 \right) L \left( d_{l0}^0 \right)}} > \tau | R^0_{00} = x \right\}} \nonumber \\
 & \stackrel{(a)}{=} \text{Pr}{\left\{ \frac{G_0 L \left( d_{00}^0 \right)}{ \sum_{l \in \mathcal{N}_u^0}{G_l L \left( d_{l0}^0 \right)}} > \tau | R^0_{00} = x \right\}} \nonumber \\
 & = \text{Pr} \left\{ 1 > \tau \sum_{l \in \mathcal{N}_u^0}{ \frac{G_l}{G_0} \left( \frac{d_{l0}^0}{\sqrt{x^2+h_0^2}} \right)^{-v}}\right\} \nonumber \\
 & \stackrel{(b)}{=} \text{Pr} \left\{ g > \tau \sum_{l \in \mathcal{N}_u^0}{ \frac{G_l}{G_0} \left( \frac{d_{l0}^0}{\sqrt{x^2+h_0^2}} \right)^{-v}}\right\} \nonumber \\
 & \stackrel{(c)}{\approx}  1 - E\left\{ \left[ 1 - \exp \left( \eta \tau \sum_{l \in \mathcal{N}_u^0}{\frac{G_l}{G_0} \left( \frac{d_{l0}^0}{\sqrt{x^2+h_0^2}} \right)^{-v}} \right) \right]^N \right\} \nonumber
\end{align}}}
\begin{equation}\label{SINRcaleq1X2}
= \sum_{n=1}^N \left( -1 \right)^{n+1} \binom{N}{n} E\left\{ \exp{\left( -n \eta \tau \sum_{l \in \mathcal{N}_u^0}{\frac{G_l}{G_0} \left( \frac{d_{l0}^0}{\sqrt{x^2+h_0^2}} \right)^{-v}}  \right)} \right\}
\end{equation}
where   $\eta = N\left( N! \right)^{-\frac{1}{N}}$.  $N$ is a parameter that controls the accuracy of the approximation and $R$ denotes the distance of the nearest BS to the typical user in which its probability density function (PDF) is $f_R \left( r \right) = 2\pi \lambda r \exp{\left(-\pi \lambda r^2 \right)}$. Also $h_0$ and $h$ denote the effective height related to the typical user and interfering users in other cells that contribute to pilot contamination. $(a)$ follows from the definition of $G_0, G_{\ell}$, in $(b)$ a dummy random variable $g$ distributed as a normalized gamma variable with parameter $N$ is used and the approximation in step $(c)$ is derived using a similar method as in \cite{CovandRatemmWave15}. 

The final expectation in (\ref{SINRcaleq1X2}) is calculated as

{\small{
\begin{align} \label{88888}
&  E\left\{ \exp{\left( -n \eta  \tau \sum_{l \in \mathcal{N}_u^0}{\frac{G_l}{G_0} \left( \frac{d_{l0}^0}{\sqrt{x^2+h_0^2}} \right)^{-v}}  \right)} \right\}  \nonumber \\
& = E \left\{ \prod_{l \in \mathcal{N}_u^0}{\exp{\left(- n \eta \tau \frac{G_l}{G_0} \left( \frac{d_{l0}^0}{\sqrt{x^2 + h_0^2}} \right)^{-v} \right)}} \right\} \nonumber \\
& =  E \left\{ \prod_{l \in \mathcal{N}_u^0}{E_{H_{\text{eff}}}\left\{ \exp\left( - n \eta \tau \frac{G_l}{G_0} \left( \frac{d_{l0}^0}{\sqrt{x^2 + h_0^2}} \right)^{-v} \right) \right\}} \right\} \nonumber \\
& = \exp \left(-2\pi \lambda \int_{R_e}^{\infty}{r \left( 1- F\left(h_0,r,x,n,\tau \right)\right)}dr \right)
\end{align}}}
where 
\begin{align}
F\left(h_0,r,x,n,\tau \right) = E_{h}{\left\{ \exp\left( - n \eta \tau \frac{G_l}{G_0} \left( \frac{\sqrt{r^2+h^2}}{\sqrt{x^2 + h_0^2}} \right)^{-v} \right) \right\}}.
\end{align}

In (\ref{88888}), the last equality comes from the probability generating functional (PGFL) of the PPP \cite{SGwireles09}. Finally, the coverage probability of the network is obtained as (10) at the bottom of this page.

\section{Tilt angle optimization}\label{covmax}
The coverage probability derived in (\ref{88888}), among other parameters, depends on the tilt angle of the BS antennas pattern as well as the users' height. In this section, using the above coverage probability as the objective function, we define an optimization problem to find the optimum tilt angle of the BSs' antenna pattern. This optimization problem can be written as follows
\setcounter{equation}{10}
\begin{align} \label{optima}
& \underset{\theta}{\arg} \, {\max{P_c}} \nonumber \\
& \text{s.t.} \nonumber \\
& 0\le \theta \le 90^{\circ}
\end{align}
%
where $P_c$ is given  by equation \eqref{coverage_probability}.  Unfortunately, the objective function in (\ref{optima}) is complicated and this problem can not be solved by analytical methods and a closed form solution can not be achieved.  
In next section, we solve this problem by numerical methods and the optimum tilt angle is found by a greedy search approach.

\section{Numerical Analysis}\label{simsec}

\begin{figure}
	\centering
	\includegraphics[scale = 0.64]{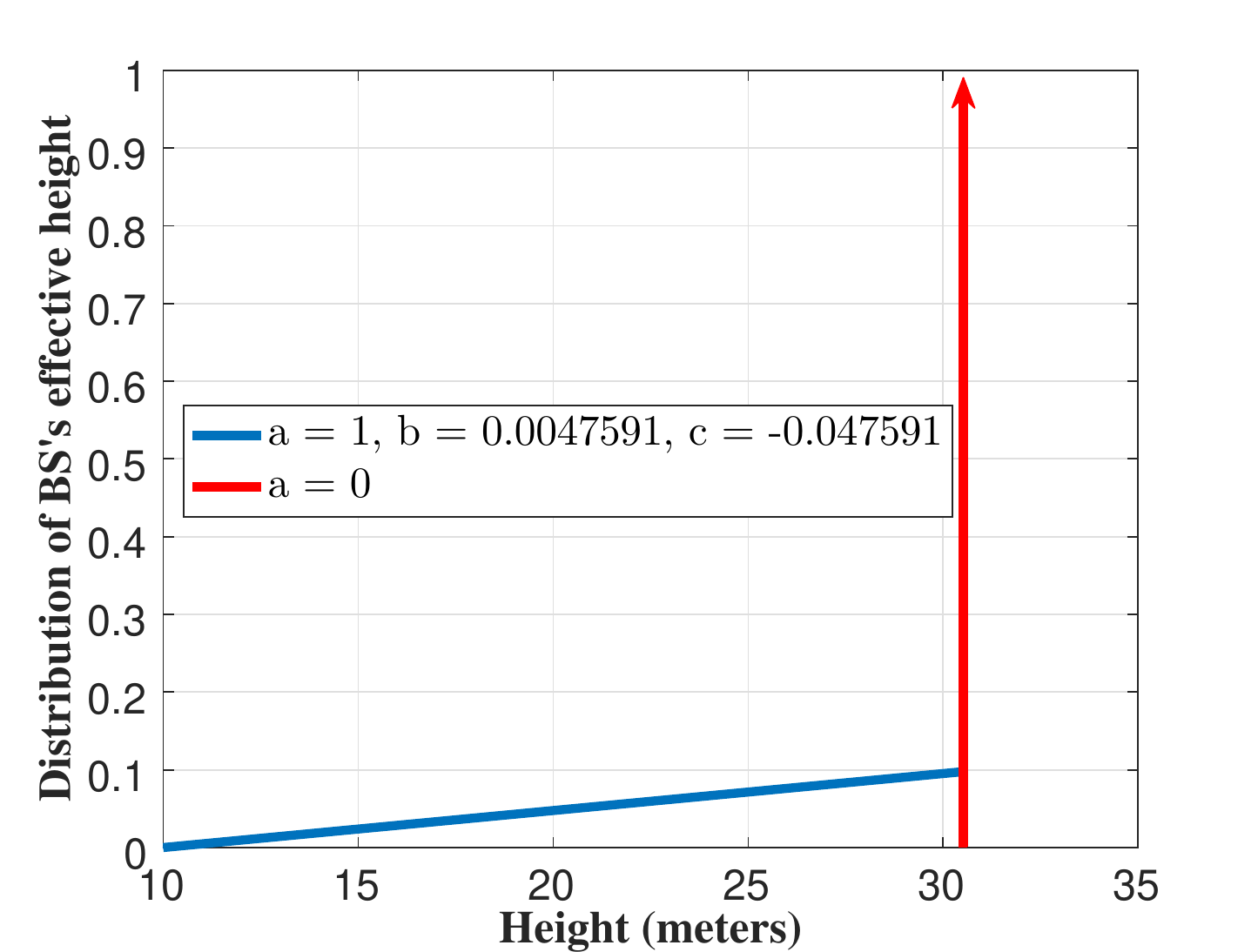}
	\caption{ Two different distributions for the BS's effective height.}
	\label{fig:arrowheight}
\end{figure}

In this section, we present the results of the computer simulations that we have performed for evaluating the performance of the proposed method. In our simulations, we study the coverage of a network which is derived in \eqref{coverage_probability}. We assume that the density of the BSs is $\lambda = 5 \times 10^{-5} $ or $ \lambda =  10^{-6} $. The typical user's height is $h_0 = 10 $ or $h_0 = 35.5$ and the path loss exponent is $v = 3.6$. Also, we assume that in equation \eqref{HeffEq} $b = 0.0047$ and $c = -0.047$. Two different case of $a = 1$ and $a = 0$ are considered which model two different scenarios of the users' height distribution. The SIR threshold ($\tau$) in the simulations is set to $4dB$ unless otherwise stated.  The resulting effective height distributions of the BSs as perceived by the users are depicted in Fig. \ref{fig:arrowheight}.

\begin{figure}
	\centering
	\includegraphics[scale = 0.6]{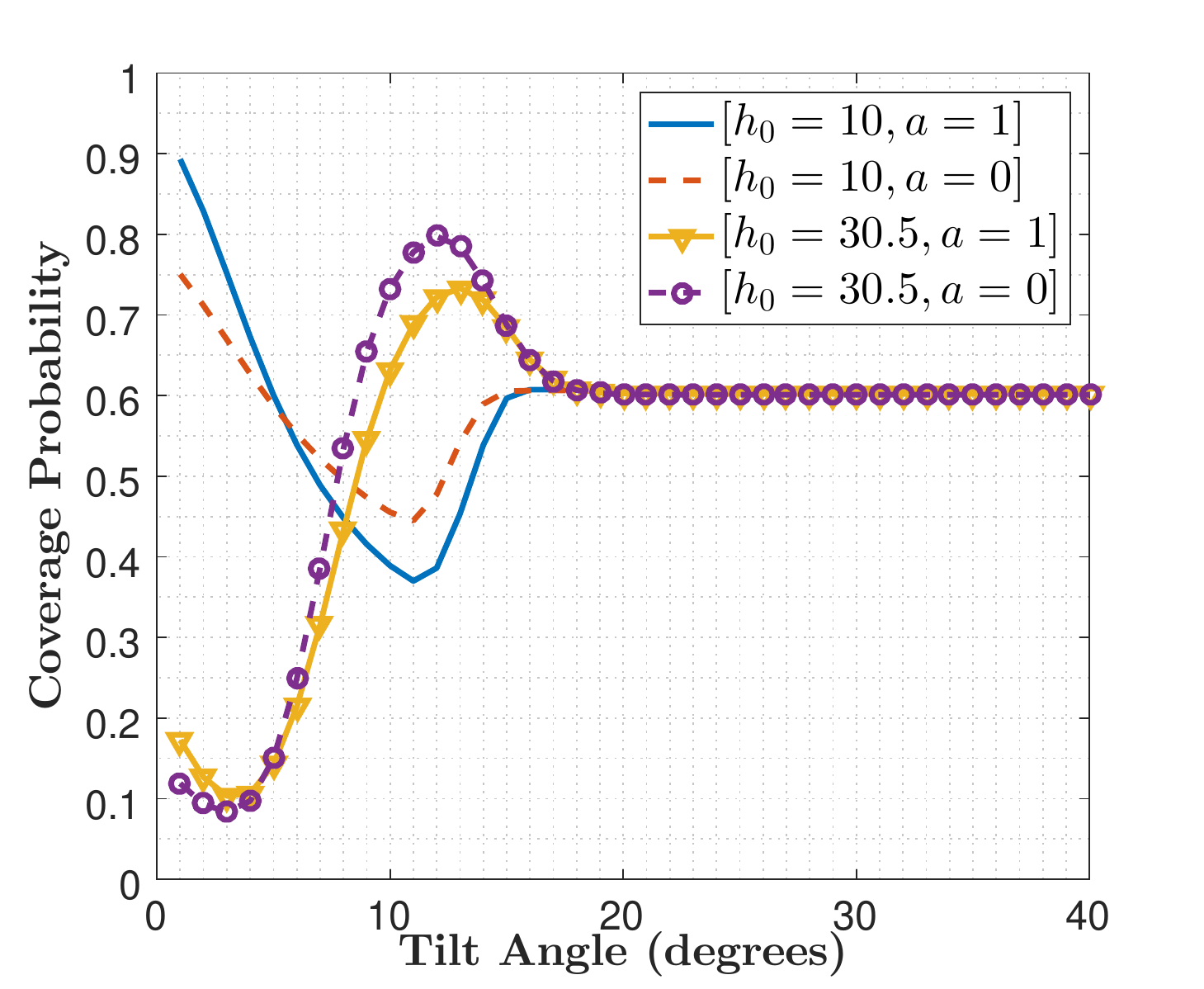}
	\caption{The coverage probability vs. tilt angle, for different cases  of users placement (in $\lambda_m= 10^{-6}$ and $\tau=4dB$).} \label{PoC_vs_tilt_R252}
\end{figure}

\begin{figure}
	\centering
	\includegraphics[scale = 0.6]{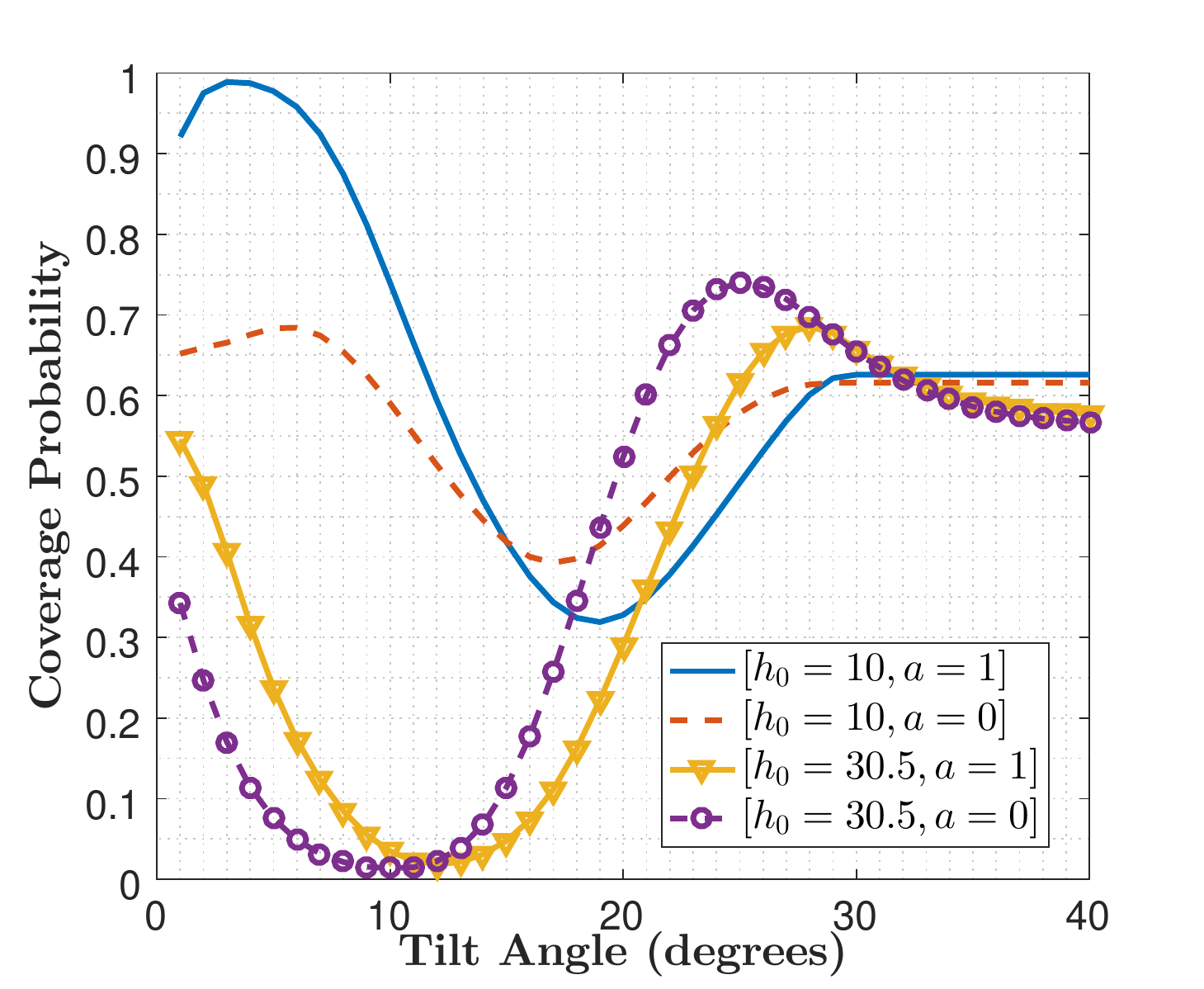}
	\caption{The coverage probability vs. tilt angle, for different cases  of users placement (in $\lambda_m=5\times10^{-5}$ and $\tau=4dB$).} \label{PoC_vs_tilt_R80}
\end{figure}

Fig. \ref{PoC_vs_tilt_R252}  represents the coverage probability of the network versus the tilt angle of the BS. In this figure, different curves are related to different users distribution and heights.  It is observed that for different users' height distribution, the optimum tilt angle varies significantly. 
The curves with $a = 0$ are related to the case that the interfering users are located on the ground and the $a = 1$ models the case that the interfering users are located in different heights above the ground. In addition, the curves with $h_0 = 30.5$ are related to the case that the typical user is on the ground and the $h_0 = 10$ models the case that the typical user is located in high altitude. The density of the BSs is assumed to be $\lambda_m=10^{-6}$ in this figure. Fig. \ref{PoC_vs_tilt_R80} shows the same results when the density of the BSs is increased to $\lambda_m = 5 \times 10^{-5}$. It is observed that by increasing the density of the BSs, the optimum tilt angle slightly increases, but the optimum tilt angle as a function of users' height distribution, shows the same behavior. In addition, the variation of the coverage probability with the tilt angle is increased.

\begin{figure}
\centering
\includegraphics[scale = 0.6]{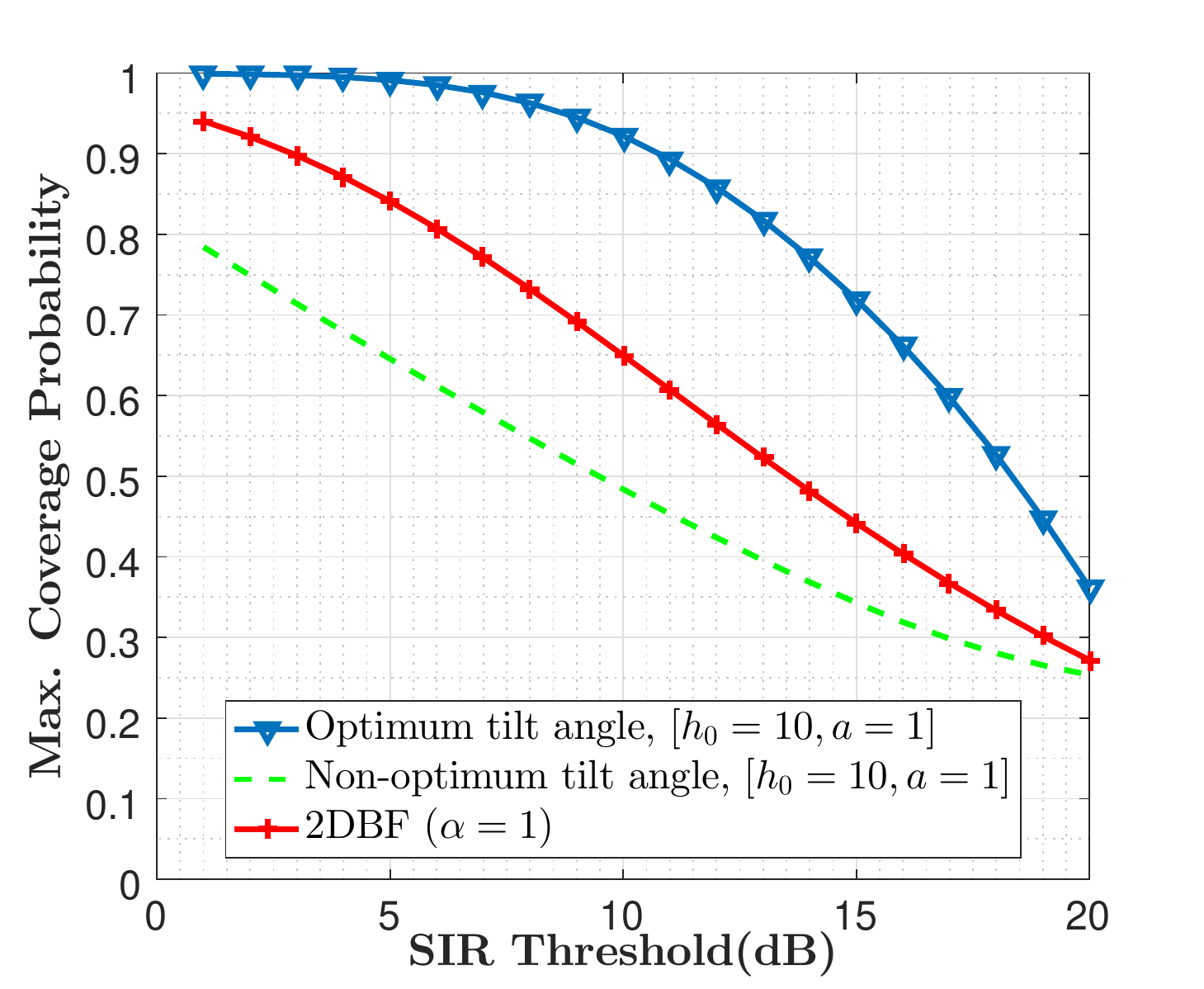}
\caption{The maximum coverage probability vs. SIR threshold for optimum and non-optimum tilt angle and its comparison with 2DBF ($\lambda_m = 10^{-6}$).} \label{PoC_vs_SINR_R252}
\end{figure}

\begin{figure}
\centering
\includegraphics[scale = 0.6]{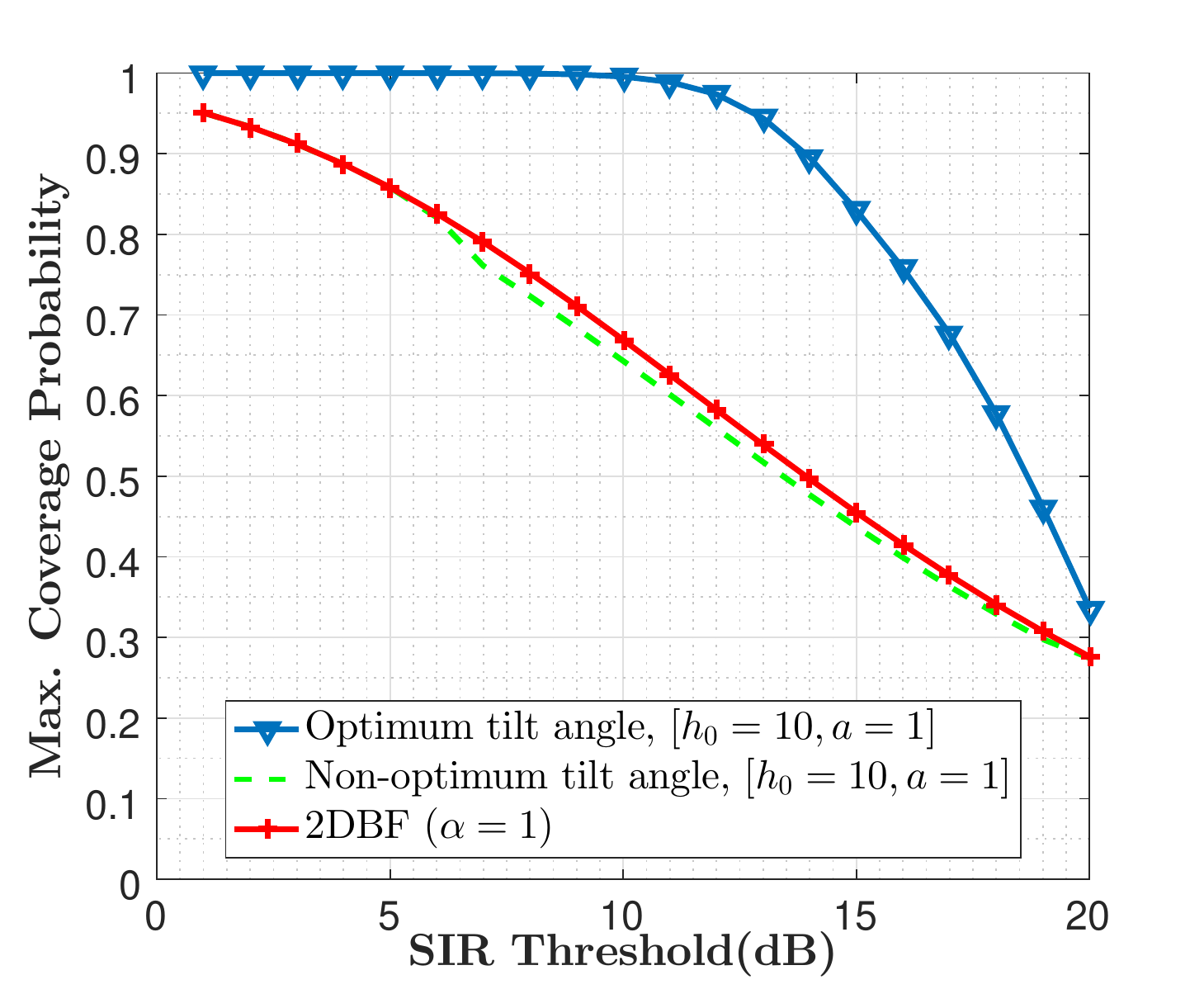}
\caption{The maximum coverage probability vs. SIR threshold for optimum and non-optimum tilt angle and its comparison with 2DBF ($\lambda_m =5\times10^{-5}$).} \label{PoC_vs_SINR_R80}
\end{figure}

Fig. \ref{PoC_vs_SINR_R252} represents the coverage probability in terms of SIR threshold. In this figure, the 'blue' curve shows the case that the height of the typical user is $h_0 = 10 $ and all the other users are distributed in different height above the ground. In this case, for each value of the SIR threshold, the optimum tilt angle is calculated and hence, the depicted coverage probabilities are the maximum values. The 'green' curve shows the case that the users have the same distribution but during finding the optimum tilt angle, it has been assumed that all users are placed on the ground. The 'red' one is also related to the 2D beamforming, omni-directional pattern in the vertical domain or $\alpha(\beta_i , \theta^i_{jk}) = 1$ in eq. \eqref{Tilt_Angle_Equation}.  The density of the BSs is assumed to be $\lambda_m=10^{-6}$ in this figure. Fig. \ref{PoC_vs_SINR_R80} shows the same results for  $\lambda_m = 5 \times 10^{-5}$. By comparing the two figures \ref{PoC_vs_SINR_R252} and \ref{PoC_vs_SINR_R80}, we can observe that by increasing the density of BS's and hence shrinking the average cell radius, advantage of obtaining optimal tilt angle by including height in calculations increases. It is obvious from Fig. \ref{PoC_vs_SINR_R80} that the blue curve has a larger difference with red curve compared to the value of difference in \ref{PoC_vs_SINR_R252} Fig.

\section{Conclusion} \label{concsec}
In this paper, we have investigated the effect of users' height on the coverage probability of a 3D beamforming-enabled multi-cell massive MIMO network. By deriving the coverage probability of the network in the asymptotic case of massive MIMO and using a stochastic geometry approach, the optimum tilt angle of the BS antenna is found to maximize the coverage probability for various 3D user distributions. Then these optimum tilt angles are compared with the 2D case and also with the 3D beamforming case when the heights of the users are ignored. It is observed when the height of the users is taken into account, using the resulting tilt angle results in a substantially larger coverage probability.

\bibliographystyle{IEEEtran}
\bibliography{Referencesbaianifar}

\end{document}